\documentclass[%
 reprint,
 amsmath,amssymb,
 aps,
 pra,
]{revtex4-1}

\usepackage{graphicx}% Include figure files
\usepackage{dcolumn}% Align table columns on decimal point
\usepackage{bm}% bold math
\usepackage{physics}
\usepackage{float}
\begin{document}

\title{Chaotic Dynamics of Inner Ear Hair Cells}
\author{Justin Faber$^1$}
\author{Dolores Bozovic$^{1, 2}$}
\affiliation{ $^1$Department of Physics \& Astronomy and $^2$California NanoSystems Institute, University of California, Los Angeles, California 90095, USA\\}
\date{\today}

\begin{abstract}
Experimental records of active bundle motility are used to demonstrate the presence of a low-dimensional chaotic attractor in hair cell dynamics.  Dimensionality tests from dynamic systems theory are applied to estimate the number of independent variables sufficient for modeling the hair cell response.  Poincar\'e maps are constructed to observe a quasiperiodic transition from chaos to order with increasing amplitudes of mechanical forcing. The onset of this transition is accompanied by a reduction of Kolmogorov entropy in the system and an increase in mutual information between the stimulus and the hair bundle, indicative of signal detection.  A simple theoretical model is used to describe the observed chaotic dynamics. The model exhibits an enhancement of sensitivity to weak stimuli when the system is poised in the chaotic regime.  We propose that chaos may play a role in the hair cell's ability to detect low-amplitude sounds.
\end{abstract}

\maketitle
%\tableofcontents

\section{Introduction}
The auditory system exhibits remarkable sensitivity, for it is capable of detecting sounds that elicit motions in the $\AA$ regime, below the stochastic noise levels in the inner ear \cite{Hudspeth14}. Fundamental processes that enable this sensitivity have still not been fully explained, and physics of hearing remains an active area of research \cite{Hudspeth}. 

Mechanical detection is performed by hair cells, which are specialized sensory cells essential for the hearing process.  They are named after the organelle that protrudes from their apical surface, and which consists of rod-like stereovilli that are organized in interconnected rows.  Incoming sound waves pivot these sterovilli, modulating the open probability of mechanically sensitive ion channels, and thus transforming motion into ionic currents into the cell \cite{LeMasurier,Vollrath}. In addition, hair cells of several species exhibit oscillations of the hair bundle, in the absence of a stimulus \cite{Benser,Martin03}. These oscillations were shown to violate the fluctuation dissipation theorem and are therefore indicative of an underlying active mechanism \cite{Martin01, Hudspeth08}. The innate motility has been proposed to play a role in amplifying incoming signals, thus aiding in the sensitivity of detection. While their role \textit{in vivo} has not been fully established, spontaneous oscillations constitute an important signature of the active processes operant in a hair cell, and provide an experimental probe for studying the underlying biophysical mechanisms \cite{Martin03}.  

The dynamics of an active bundle have been described using the normal form equation for the Hopf bifurcation \cite{Eguiluz00, Kern}. Several studies have furthermore proposed that a feedback process acts on an internal control parameter of the cell, tuning it toward or away from criticality \cite{Camalet00,Shlomovitz}. With the inclusion of dynamic feedback, the theoretical models required three state variables, a dimension that is sufficient to support a chaotic regime, according to the Poincar\'e-Bendixson theorem \cite{Poincare}. Numerical simulations indeed predicted a small positive Lyapunov exponent, indicative of weak chaos in the innate bundle motion \cite{Shlomovitz}.  Another numerical study that explored an 12-dimensional model of hair cell dynamics showed the presence of chaos and proposed that the sensitivity of detection to very low-frequency stimuli would be optimal in a chaotic regime \cite{Neiman}.

The presence of chaos may help to explain the extreme sensitivity of hearing, as it has been shown in nonlinear dynamics theory that chaotic systems can be highly sensitive to weak perturbations \cite{Pikovsky, Glass}. In the present manuscript, we therefore explore experimentally whether innate bundle motility exhibits signatures of chaos \cite{Strogatz}.  Since establishing the dimensionality of the system is crucial for accurate modeling of this remarkable mechanical detector, we apply a dimensionality test to estimate the number of state variables required to describe the dynamics of an auditory hair cell. Further, we examine the effect of an applied signal on the chaoticity of bundle motion. For this purpose, we construct Poincar\'e maps of the oscillator, subject to varying amplitudes of external forcing. We quantify the degree of chaos by computing the Kolmogorov entropy associated with the spontaneous and driven oscillation of a hair bundle. As a measure of the sensitivity to external perturbation, we compute the mutual information extracted from the applied signal by the oscillatory bundle.  Finally, we present a simple theoretical model that reproduces the quasiperiodic and chaotic dynamics that were observed experimentally.  We use the theoretical model to demonstrate that a system poised in the chaotic regime shows an enhanced sensitivity to weak stimuli.

\begin{figure}[h!]
\includegraphics[width=85mm]{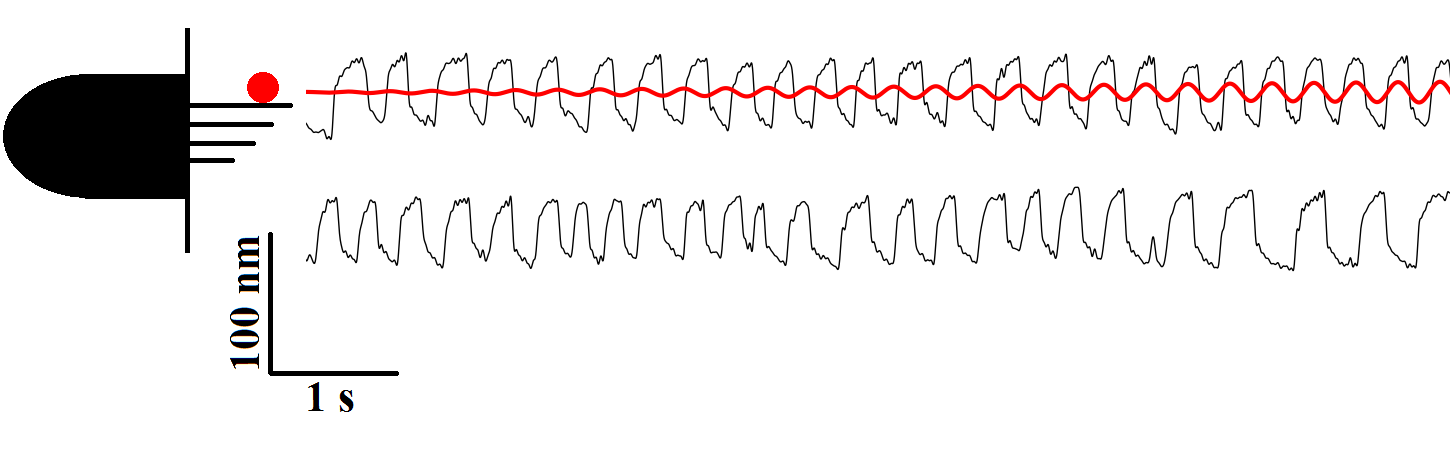}
\caption{Position of a driven hair bundle (top) and spontaneously oscillating hair bundle (bottom).  The positive direction corresponds to the direction of channel opening.  red (thick) and black traces correspond to the stimulus waveform and the hair bundle response, respectively. The schematic image of a hair cell (left) describes the bundle of stereovilli protruding from the cell body. The red dot depicts the location of probe attachment to the hair bundle.}
\label{fig:hair_bundle}
\end{figure}

\section{Experimental Methods}
Experiments were performed \textit{in vitro} on hair cells of the amphibian sacculus, an organ specializing in low-frequency air- and ground-borne vibrations. Sacculi were excised from the inner ears of North American bullfrogs (\textit{Rana catesbeiana}), and mounted in a two-compartment chamber \cite{Benser}.  Spontaneously oscillating hair bundles were accessed after digestion and removal of the overlying otolithic membrane \cite{Martin01}.  All protocols for animal care and euthanasia were approved by the UCLA Chancellor’s Animal Research Committee in accordance with federal and state regulations.  To deliver a stimulus to the hair bundles, we used glass capillaries that had been pulled with a micropipette puller.  These elastic probes were treated with a charged polymer that improves adhesion to the hair bundle.  Innate oscillations persisted after the attachment of probes with stiffness coefficients of $\sim$ 100 $\mu$N/m. The piezoelectric actuator was controlled with LabVIEW to apply sinusoidal stimulation of varying amplitudes for selected constant frequencies.  Hair bundle motion was recorded with a high-speed camera at frame rates of 500 Hz or 1 kHz.  The records were analyzed in MATLAB, using a center-of-pixel-intensity technique to determine the mean bundle position in each frame.  Typical noise floors of this technique, combined with stochastic fluctuations of bundle position in the fluid, were 3 -- 5 nm. FIG \ref{fig:hair_bundle} shows representative traces of active bundle motion, subject to mechanical forcing.

\section{Dimensionality Test}
A useful technique for estimating the dynamical dimension, $d_L$,  of a time series relies on the reconstruction of the phase space using delayed coordinates. It has been shown that this delayed-coordinate map from the original $d_L$-dimensional smooth compact manifold M to  $\mathbb{R}^{d_E}$  is diffeomorphic, provided that ${d_E} > 2{d_A}$, where ${d_A}$ is the box-counting dimension of the original attractor, and ${d_E}$ is the embedding dimension \cite{Eckmann85, Sauer, Takens}. Frequently, a lower embedding dimension, $d_L$, is sufficient to fully unfold the attractor, but it is necessary that $d_E \geq d_L$ \cite{Abarbanel, Abarbanel93}.  In finding the optimal embedding dimension, we set an upper bound on the dimension of the original dynamics.  From the original time series $x(t)$, we construct the vector,

\begin{eqnarray}
\vec{X} = [x(t), x(t + \tau), x(t + 2\tau), ..., x(t + (d_{E}-1)\tau)],
\label{eq:two}
\end{eqnarray}

\noindent for each point in the time series; $\tau$ is chosen as the time at which the average mutual information of the series exhibits its first minimum  \cite{Fraser}.   For each point in the reconstructed phase space, we compute the unit vector,

\begin{eqnarray}
\hat{u}(t_{n}) = \frac{\vec{X}(t_{n+1}) - \vec{X}(t_{n})}{||\vec{X}(t_{n+1}) - \vec{X}(t_{n})||}
\label{eq:three},
\end{eqnarray}

\begin{figure}[h!]
\includegraphics[width=85mm]{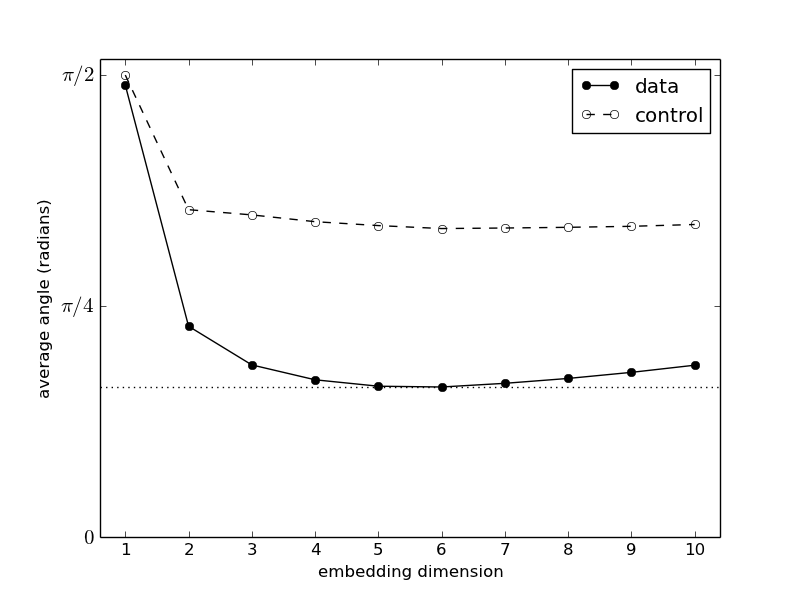}%
\caption{\label{fig:vector_angle} Average angle between neighboring flow vectors as the embedding dimension is varied. The angles are calculated for a 1 minute recording of a spontaneously oscillating hair bundle, obtained at 1 kHz sample rate.  Data was filtered with a low-pass filter to remove high-frequency stochastic processes. The cutoff frequency was set to 100 Hz, sufficiently above the dominate frequency of the hair bundle ($\sim$20 Hz).}
\end{figure}

\noindent where $t_n$ is the time of the $n^{th}$ observation.  These unit vectors point in the direction of local flow on the attractor.  For deterministic systems, neighboring unit vectors are nearly parallel if the attractor is densely sampled, and if the time series is not contaminated by stochastic processes. The flow therefore becomes smooth when the embedding dimension is high enough to fully unfold the attractor. The smoothness of this reconstructed phase space can be quantified by finding the average angle that each unit vector makes with its nearest neighbor \cite{Abarbanel, Kaplan}. Starting with one embedding dimension, we calculate the average angle among all unit vectors, then increase the embedding dimension, and repeat the calculation.  The average angle is minimized when using the optimal embedding dimension. 

For a spontaneously oscillating hair bundle, the phase space fully unfolds between three and six dimensions (FIG \ref{fig:vector_angle}). To verify these results, we perform the same analysis on a surrogate data set, generated from the original data.  We multiply each Fourier components by a random phase, creating a stochastic signal with the power spectrum and the autocorrelation function identical to those of the original data set.  The surrogate data set does not yield a minimum similar to the original data, and the flow along phase space trajectories is much less smooth.  These results indicate that, although stochastic processes are present in our system, there is an underlying low-dimensional attractor. Further, up to six differential equations should be sufficient to describe the dynamic behavior of an active hair cell bundle.

\section{Poincar\'e Maps}
\begin{figure}[h!]
   \includegraphics[width=85mm]{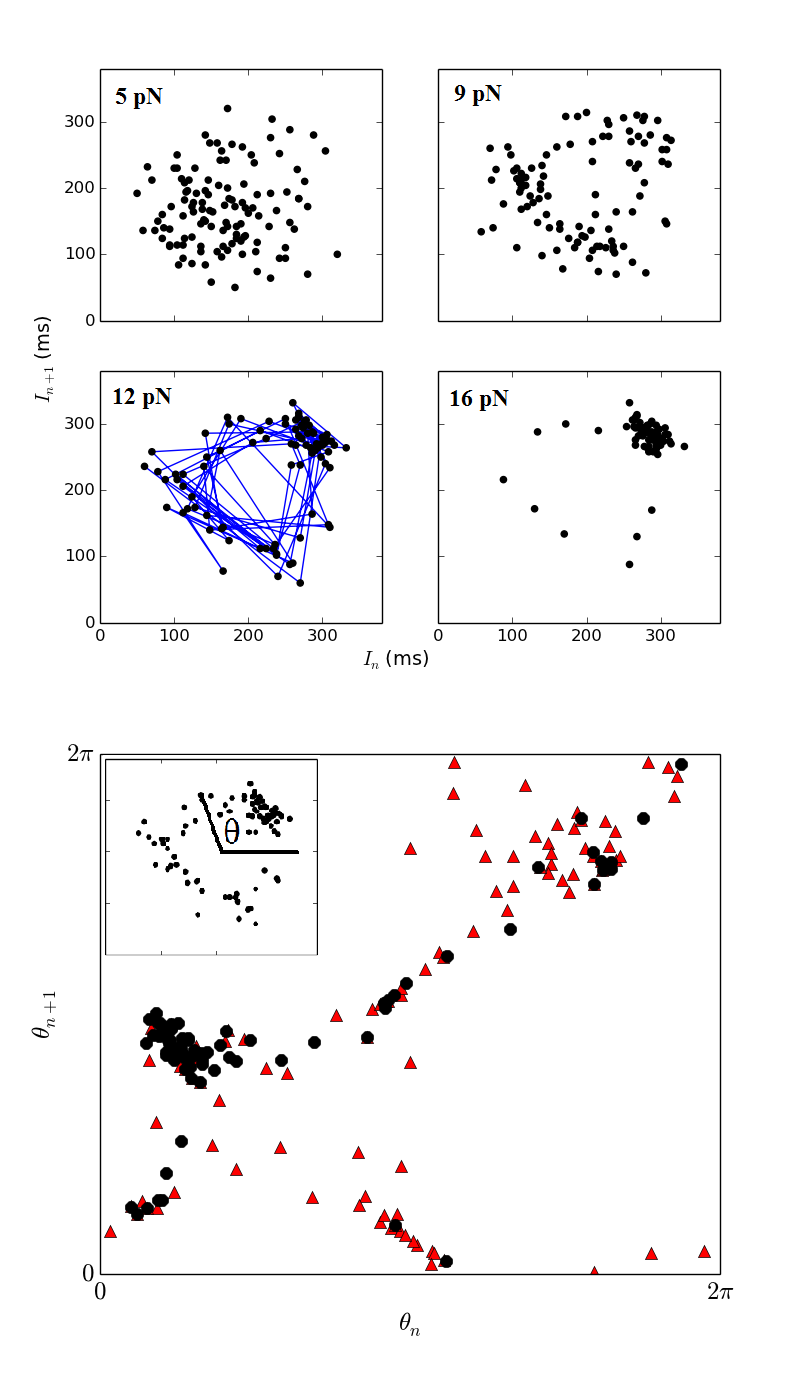}
\caption{\label{fig:poincare} Top: Poincar\'e maps representing oscillations of hair bundles driven by a sinusoidal stimulus below the resonance frequency ($\omega \sim \frac{2}{3} \omega_0$).  Blue lines connect consecutive points in the series.  Bottom:  Angles formed by vectors from the center to each point in the Poincar\'e map (inset).  Red triangles and black points represent data obtained for stimulus amplitudes of 5 pN and 13 pN, respectively.}
\end{figure}

The Poincar\'e map provides a powerful tool for observing the dynamics of a nonlinear system in a lower dimensional space.  For a perfectly periodic signal, the map takes the form of a single point, while for a quasiperiodic signal the map comprises a ring-like structure.  This ring represents a cross section of the torus on which the trajectories of the phase space are confined.  The occurrence of trajectories that fall off the surface of this torus indicates a quasiperiodic transition to chaos via torus breakdown \cite{Ott, Garfinkel}.  Stochastic and high-dimensional processes yield a cloud-like Poincar\'e map that has no internal structure.  Poincar\'e maps constructed for the motion of a hair bundle subject to sinusoidal stimuli at varying amplitudes of forcing are shown in FIG \ref{fig:poincare}.  To construct these Poincar\'e maps from our recordings, we determine the discrete time series, $[I_n]$, where subsequent elements are the time intervals between the steepest rising flanks of consecutive bundle oscillations.  We then plotted the $n^{th}$ versus the $(n+1)^{th}$ point of the series to yield the Poincar\'e map.  As the series constitutes an observable in phase space, embedding theory can be applied.

At low stimulus amplitudes, the Poincar\'e maps form a cloud-like structure.  At higher stimulus amplitudes, a ring structure emerges from the cloud. Consecutive points within the sequence migrate around the edge of the ring, rather than crossing over the center, indicative of a quasiperiodic behavior.  When the stimulus amplitude is increased above approximately 15pN, the hair bundle follows the stimulus, causing the ring structure to collapse onto a point.  This quasiperiodic transition was observed only when the stimulus frequency was below the resonance frequency of the hair bundle. 

As an additional test for the presence of low-dimensional chaos, a line is drawn from the center of the ring to each point in the sequence, and the angle formed by these lines and the abscissa is computed.  This series of angles yields a new map, $\theta_{n+1} = f(\theta_n)$.  When chaos arises via a quasiperiodic transition, points fall off the surface of a 2-torus, since chaotic dynamic can be described by no fewer than three state variables.  As a result of the torus breakdown, the map $f$ becomes noninvertible and ceases to be a function \cite{Garfinkel}. As seen in FIG \ref{fig:poincare}, the map is non-invertible for a weak stimulus. It approaches an invertible map when a stronger stimulus is applied to the bundle, indicating the disappearance of low-dimensional chaos.  For stimulus frequencies near the hair bundle’s natural frequency, the Poincar\'e maps transition directly from a cloud to a point, bypassing quasiperiodic dynamics.  For stimulus frequencies above the hair bundle's resonance frequency, the hair bundle exhibits a flicker between 1:1 and 2:1 mode-locking, over a range of forcing amplitudes (see supplemental material).

\section{Complexity and Entropy}

An additional test for the presence of low-dimensional chaos in a nonlinear system can be obtained from measurements of the permutation entropy and statistical complexity \cite{Martin, Bandt, Rosso, Gekelman}. Beginning with a time series, [$x_1$,...,$x_N$], we take sub-chains of length d, ([$x_i$,...,$x_{i+d-1}$]). There are $d!$ possible permutations of amplitude ordering within the sub-chain (d! different states). A data set of length N produces $N-(d-1)$ sub-chains. The probability distribution, P, of these states is used to calculate the normalized Shannon entropy, $H(P)$,

\begin{eqnarray}
S(P) = -\sum_{i=1}^{i=d!}p_i\ln(p_i)
\label{eq:S}
\end{eqnarray}

\begin{eqnarray}
H(P) = \frac{S(P)}{\max(S)} = \frac{S(P)}{\ln(N)}
\label{eq:H}
\end{eqnarray}

and the Jensen-Shannon complexity,

\begin{eqnarray}
C_{js} = -2\frac{S(\frac{P+P_e}{2}) - \frac{1}{2}S(P) - \frac{1}{2}S(P_e)}{\frac{N+1}{N}\ln(N+1) - 2\ln(2N) + \ln(N)}H(P)
\label{eq:complexity}
\end{eqnarray}

\noindent where $P_e$ is the probability distribution of maximum entropy (with all states equally probable). 

\begin{figure}[h!]
\includegraphics[width=85mm]{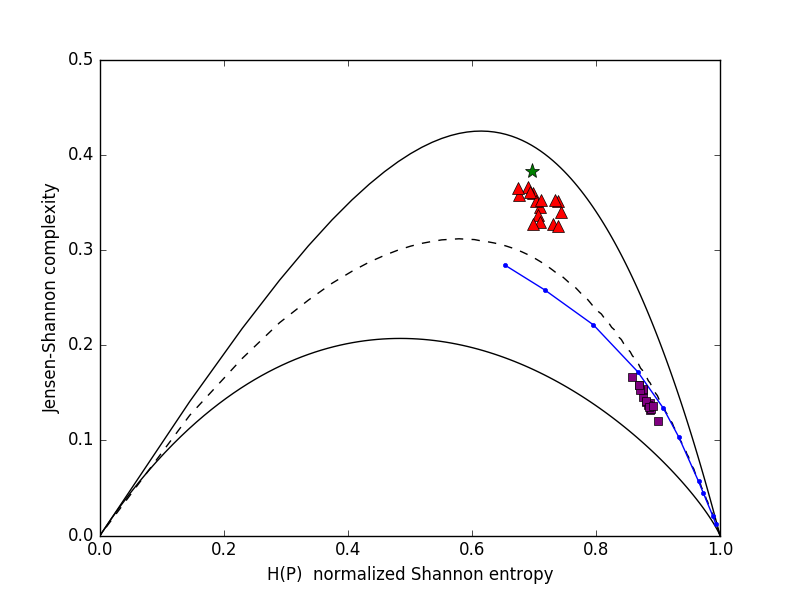}%
\caption{\label{fig:Comp_Ent} Complexity-entropy diagram for a spontaneously oscillating hair bundle fully sampled (purple squares) and sub-sampled by selecting every fifth point (red triangles).  The green star represents the theoretical model, with the system poised in the chaotic regime.  The blue points correspond to fractional Brownian motion with Hurst exponent ranging from 0.02 to 0.98.  The black curves confine all possible probability distributions.  The dashed curve is the half-way point between the upper and lower bounds and serves as an approximate boundary between stochastic and chaotic probability distributions.}
\end{figure}

All possible probability distributions are confined to a specific region in the complexity-entropy plane. Jointly, the two measurements allow one to determine whether a chaotic attractor or stochastic noise dominates the dynamics of a system. The lower part of the complexity-entropy region (low-complexity) is occupied by probability distributions of stochastic signals, while the high-complexity region is occupied by probability distributions of signals with low-dimensional chaos.  We apply this test to our measurements of spontaneously oscillating hair bundles (FIG \ref{fig:Comp_Ent}). We selected d=5, which yields 120 (i.e. 5!) possible states. Similar results were obtained for d=4 and d=6.

Sub-sampling is a useful technique for extracting a low-dimensional chaotic process from a signal contaminated by noise. A purely stochastic process is hardly affected by sub-sampling; if a chaotic attractor is present in the system, however, it may emerge in the sub-sampled data \cite{Maggs}.  Recordings of hair bundle oscillations were sampled at 1 kHz, resulting in a Nyquist frequency of 500 Hz.  Sub-sampling every fifth point reduces the Nyquist frequency to 100 Hz, sufficiently above the dominant frequency in the signal ($\sim$20 Hz). The sub-sampled data set yields statistical features of low-dimensional chaos (FIG \ref{fig:Comp_Ent}).

\section{Kolmogorov Entropy}
Kolmogorov entropy quantifies the level of chaos in the trajectory of a dynamical system \cite{Kolmogorov}. Given knowledge of the state of a system at a particular time, obtained with some uncertainty, Kolmogorov entropy (K-entropy) measures how well one can predict its state at a later time. K-entropy therefore provides a measure of how rapidly neighboring trajectories diverge, and reflects the rate at which the system is producing information. Limit cycles produce no information: for a given observation of the state of the system, all future states can be predicted with the same uncertainty as the original measurement. In contrast, systems exhibiting either low- or high-dimensional chaos constantly produce information.  Hence, K-entropy is zero for limit cycles, positive for systems with low-dimensional chaos or noise, and infinite for purely stochastic processes.  

To calculate the K-entropy, we divide the reconstructed phase space into hypercubes, and select a starting hypercube to be one that contains many points of the trajectory. We track the trajectories emerging from these points to calculate a time-dependent probability distribution over the hypercubes.  The resulting distribution yields the Shannon entropy as a function of time; its time-averaged derivative is defined to be the K-entropy \cite{Kolmogorov, Kim}.  An embedding dimension of five was chosen for this analysis; nearly identical results were obtained with four or six embedding dimensions.  The choice of the number of bins per dimension did not affect the results of this analysis. We used two bins per dimension, a natural choice due to the bimodal distribution in position of the spontaneously oscillating hair bundle.  

\begin{figure}[h!]
\includegraphics[width=85mm]{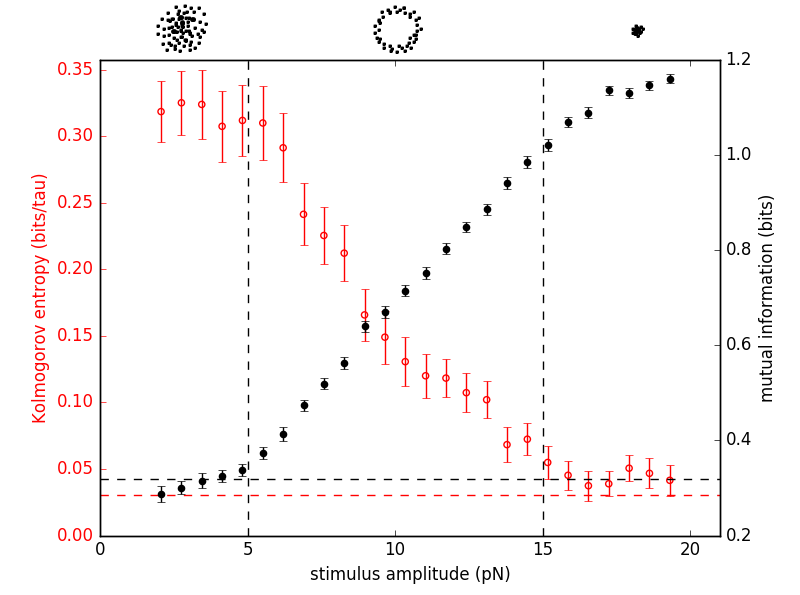}%
\caption{\label{fig:kolmogorov} 
The Kolmogorov entropy of an oscillatory hair bundle subject to a sinusoidal drive (open-red) and mutual information between the bundle and the stimulus (filled-black), plotted as a function of the stimulus amplitude.  Vertical dashed lines delineate approximate regions that correspond to Poincar\'e maps displaying cloud, ring, and point structures, as illustrated by the schematic diagrams above the graph. Noise floors are indicated by horizontal dashed lines.  The noise floor on the K-entropy is estimated by tracking the motion of a passive hair bundle driven by a 10 -- 25 pN sinusoidal stimulus.  The noise floor on the mutual information is estimated based on bundle oscillations in the absence of a stimulus.  Uncertainties in K-entropy are estimated by taking the standard deviation of the residuals from the linear fit of Shannon entropy with time.  Uncertainties in mutual information are estimated from a bootstrapping technique that incorporates the uncertainties in position measurements}
\end{figure}

We observe that the Kolmogorov entropy associated with active bundle motility is reduced by the application of a sinusoidal drive  (FIG \ref{fig:kolmogorov}).  The majority of the reduction in K-entropy occurs during the quasiperiodic transition from chaos to order, the regime in which the Poincar\'e maps produce a ring structure.  K-entropy plateaus near zero, for forcing amplitudes above $\sim$15 pN. We note that a noiseless system would plateau exactly at zero; the finite value of the plateau is due to noise inherent in the experimental recording.

Detection by an individual hair cell is quantified by calculating the mutual information between the stimulus signal and the receiver response.  Mutual information between two processes X and Y is defined as

\begin{eqnarray}
M_{XY} = \sum_{x\in X}\sum_{y\in Y}P(x,y) \log(\frac{P(x,y)}{P(x)P(y)})
\label{eq:four}
\end{eqnarray}

\noindent where $P(x, y)$ is the probability of simultaneously finding values x and y from probability distributions $P(x)$ and $P(y)$, respectively.  To calculate the $P(x)$, we construct histograms of the stimulus and bundle positions using 50 bins; qualitatively similar results are obtained using 5, 10, 20 and 100 bins.  The mutual information between the applied stimulus and the hair bundle oscillation rises  above zero, for signals of $\sim$5 pN. The observed detection threshold is even lower when a stimulus is applied at the resonance frequency of the cell (see supplemental material). The mutual information exhibits the steepest increase with amplitude of the applied force in the quasiperiodic regime.

\section{Theoretical Model}
To capture the chaotic dynamics of an oscillatory hair bundle, we apply a variant of a previously proposed model \cite{Shlomovitz}, which was shown to account for the complex multi-mode hair bundle oscillations. It was shown to reproduce the observed suppression and recovery of oscillation, following strong mechanical forcing.  Further, it exhibits multi-mode locking to the stimulus for a wide range of driving frequencies, in agreement with experimental results. The model consists of three dynamical variables, the minimum dimension capable of supporting a chaotic regime.  

The transition between oscillatory and quiescent states is described using the normal form equation of the Hopf bifurcation, 

\begin{eqnarray}
\frac{dz}{dt} = z(\mu - i\omega_0 + A|z|^2 - B|z|^4) + f_{A}\cos(w_dt)
\label{eq:one},
\end{eqnarray}

\noindent where $z(t)$ describes the dynamic state of the bundle with the real part, $x(t)$, representing the bundle position. $f_{A}$, $\omega_{d}$, and $\omega_{0}$ denote the driving force, driving frequency, and natural frequency, respectively. 

The control parameter, $\mu$, is associated with the probability of the system being in the oscillatory or the quiescent state.  We assume the parameter to be a dynamic variable, with its rate of change described by

\begin{eqnarray}
\frac{d\mu}{dt} = k_{on} - k_{off}\Theta(x - x_0) + \alpha f_{A}
\label{eq:two},
\end{eqnarray}

\noindent where $k_{on}$ and $k_{off}$ are rate constants.  The Heaviside step function, $\Theta$, serves as an approximation of the Boltzmann distribution related to the opening probability of the ion channels \cite{Hudspeth08}.  We introduced the $\alpha f_{A}$ term to the original model, to capture the entrainment of the bundle by a strong stimulus.

We demonstrate that numerical simulations based on this model reproduce well the characteristic features of the experimental results.  Sinusoidal stimuli of linearly increasing amplitude elicit a quasiperiodic transition from chaos to order.  This transition is accompanied by a reduction in the Kolmogorov entropy and an increase in the mutual information between the stimulus and the response, indicative of signal detection (see supplemental material).  The model also produces dynamics that lie within the same region of the complexity-entropy plane as those of the experimental recordings (FIG \ref{fig:Comp_Ent}).

We next explore the effects of chaoticity on the sensitivity to a weak stimulus.  Chaoticity is most easilty quantified by the largest Lyapunov exponent \cite{Strogatz}. In the absence of a stimulus, Lyapunov exponents were calculated by tracking the rate of divergence of neighboring trajectories in the 3-dimensional phase space.  To compute the sensitivity, we use a scaled version of the traditional linear response function \cite{Martin01}: 

\begin{eqnarray}
\bar{\chi}(w_d) = \frac{|<\tilde{x}(w_d)> - <\tilde{x_0}(w_d)>|}{<\tilde{x_0}(w_d)>},
\label{eq:chi}
\end{eqnarray}

\noindent where $<\tilde{x}(w_d)>$ and $<\tilde{x_0}(w_d)>$ are the Fourier components of the bundle position at the stimulus frequency in the presence and absence of stimulus, respectively.  

\begin{figure}[h!]
\includegraphics[width=85mm]{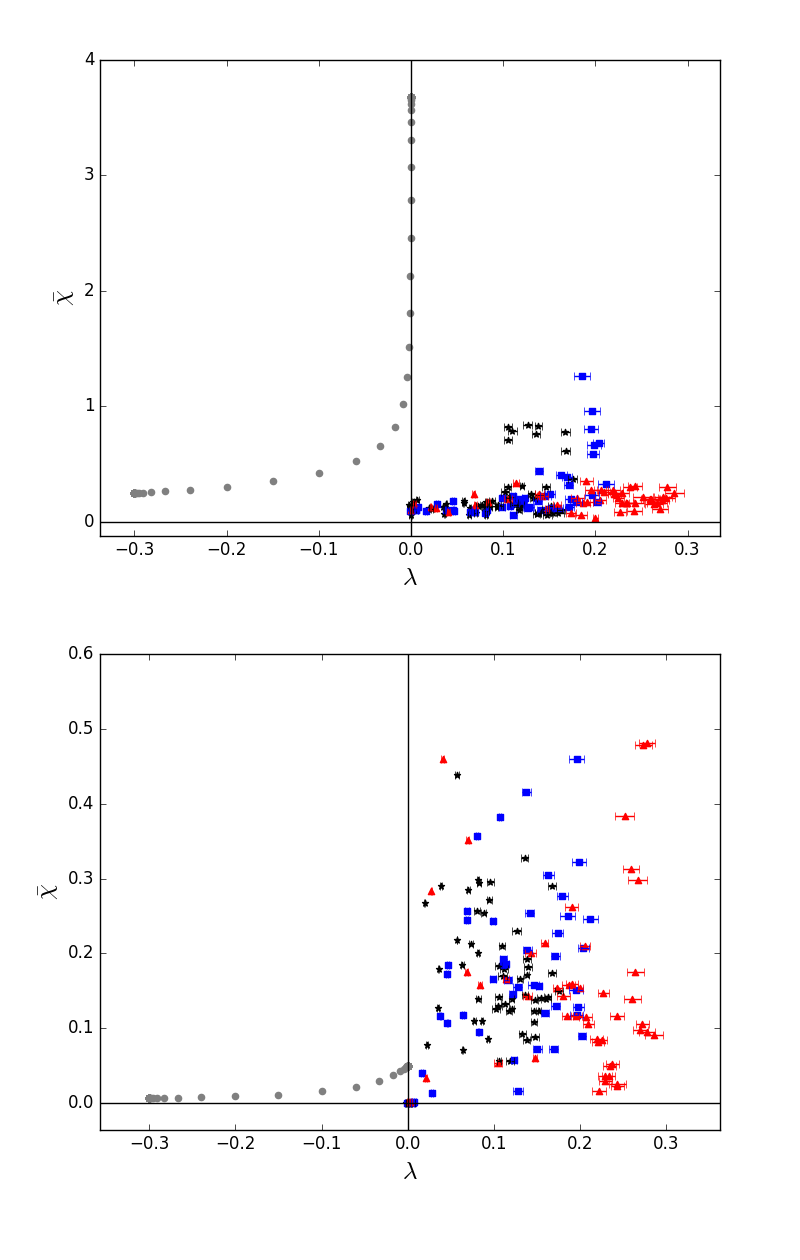}
\caption{\label{fig:chi_lam} 
Scatter plot of the largest Lyapunov exponent ($\lambda$) versus the scaled sensitivity for $f_A = 0.2$ (top) and $f_A = 0.002$ (bottom).  These numerical values correspond to forcing amplitues of approximately 1 pN and 0.01 pN, or probe displacements of 10 nm and $1 \AA$, respectively.  Blue squares, red triangles and black stars correspond to variations in parameters $k_{on}$, $k_{off}$ and $x_0$, respectively.  Grey points represent variations in $A$ when the system is in the limit cycle regime.  In this regime, the Lyapunov exponent of largest magnitude is calculated analytically.}
\end{figure}

We varied the parameters of the model, to obtain a broad range of Lyapunov exponents. For each set of parameters, we imposed weak, moderate, and strong stimuli, and calculated the scaled sensitivity. For moderate and strong forcing, the highest sensitivity is observed when the system is not chaotic, with the Lyapunove exponent negative, but approaching zero. However, for a weak stimulus ($f_A < 0.1$), the model shows a significantly greater sensitivity when poised in the chaotic regime (FIG \ref{fig:chi_lam}).

\section{Conclusions}
The auditory system has provided an experimental testing ground for theoretical work on nonlinear dynamics \cite{Eguiluz00, Camalet00}, nonequilibrium thermodynamics \cite{Dinis}, and condensed matter theory \cite{Risler}. The fundamental questions on hearing pertain to its sensitivity, frequency selectivity, rapidity of detection, and the role of an active mechanism in shaping the response. Models based on dynamic systems theory have successfully described a number of empirical phenomena. However, the theoretical models have mostly focused on stable dynamics, exploring either the limit cycle regime, or the quiescent regime in the vicinity of a bifurcation. 

Our results provide an experimental demonstration that a low-dimensional chaotic attractor arises in the dynamics of active hair bundle motility. Ring-like structures in the Poincar\'e maps, positive Kolmogorov entropy, non-integer correlation dimension (see supplemental material) and the location of the time series in the complexity-entropy plane are all indicators of low-dimensional chaos, not stochastic processes.  Because chaos can not exist in dynamical systems of dimension lower than three, the two-dimensional models extensively used in the field are insufficient for capturing the dynamics.  We estimate that three to six independent variables are needed to correctly characterize the dynamics of the system. 

Further, we find that chaos is removed by the application of a signal, with different transitions from chaos to order observed when oscillations are entrained by stimuli below, near, or above the characteristic frequency of the cell. Hair bundles have thus far been viewed as nonlinear mechanical detectors, and the linear response used to characterize their sensitivity. We propose that information theory provides a useful and complementary tool for analyzing the response of a hair cell, which can be viewed as a computational device that serves to extract information about the external acoustic environment. We hence apply an information theoretic approach to quantify the detection of a signal by an individual hair cell. The steepest dependence of information gain with stimulus amplitude is observed in the dynamic regime that corresponds to a transition from chaos to order. Hair bundles poised in the chaotic regime exhibit measurable increases in mutual information even at pN levels of stimulus, indicating that chaotic dynamics of innate motility are consistent with high sensitivity of detection. 

Our theoretical model, which includes a feedback equation for the control parameter of the system, describes well the dynamics observed experimentally. When poised in the chaotic regime, the system shows an enhanced sensitivity to weak stimulus. We therefore propose that hair cells of the auditory system harness the presence of chaos to achieve high sensitivity. We hypothesize that chaotic dynamics may be a ubiquitous feature of nonlinear biological systems, which typically exhibit many degrees of freedom. It is therefore important to understand the impact of chaos on the sensory perception in living systems. Future work entails developing experimental techniques for modulating the degree of chaos in bundle dynamics, to assess the impact of chaoticity on the sensitivity of detection. Subsequent theoretical studies entail including noise terms in the model, to explore the interaction between stochastic processes and deterministic dynamics.

\begin{acknowledgments}
The authors gratefully acknowledge support of NIDCD, under grant R21DC015035, and NSF, under grant IOS-1257817. The authors thank Dr. Sebastiaan Meenderink for developing the software used for tracking hair bundle movement.  The authors thank Dr. Alan Garfinkel for helpful discussions on numerical methods.
\end{acknowledgments}

\section{Supplemental Material}

\subsection{Correlation Dimension}

The fractal dimension of an attractor reflects the space filling capacity of its trajectories.  The correlation dimension provides a similar measure and is frequently used to estimate the fractal dimension of a system that is contaminated by noise \cite{Grassberger83, Ben-Mizrachi}.  The correlation dimension can never exceed the number of degrees of freedom of the dynamical system, and hence yields a lower bound.  To measure the correlation dimension, the phase space is reconstructed using the delayed-coordinate technique (see section on dimensionality test).  Hyperspheres are constructed that are centered on each of the phase space points. The correlation sum is defined as

\begin{eqnarray}
C(r) \equiv  \lim_{N\to\infty} \frac{1}{N^2} \sum_{i,j = 1}^{N} \Theta(r - \norm{\vec{X_i} - \vec{X_j}}),
\label{eq:C}
\end{eqnarray}

\noindent where $\Theta$ is the Heaviside step function, and $\vec{X_i}$ is the vector from the origin to the location of the $i^{th}$ point in reconstructed phase space. The correlation sum is a function of the hypersphere radius, and for small values of r, should obey the power law,

\begin{eqnarray}
C(r) \propto r^\nu,
\label{eq:four}
\end{eqnarray}

\noindent where $\nu$ is the correlation dimension. To extract the dimension $\nu$ from the data, we plot $\log(C(r))$ versus $\log(r)$ and find the slope of the linear regime.  We repeat this for an increasing number of embedding dimensions, until a plateau occurs in the values of $\nu$.  This plateau onset is expected to occur when the embedding dimension exceeds the correlation dimension \cite{Ding}.  This plateau never occurs for a time series dominated by stochastic processes.  Integer values of $\nu$ imply a non-chaotic attractor while non-integer values of $\nu$ are indicative of a chaotic attractor.

\begin{figure}[h!]
\includegraphics[width=85mm]{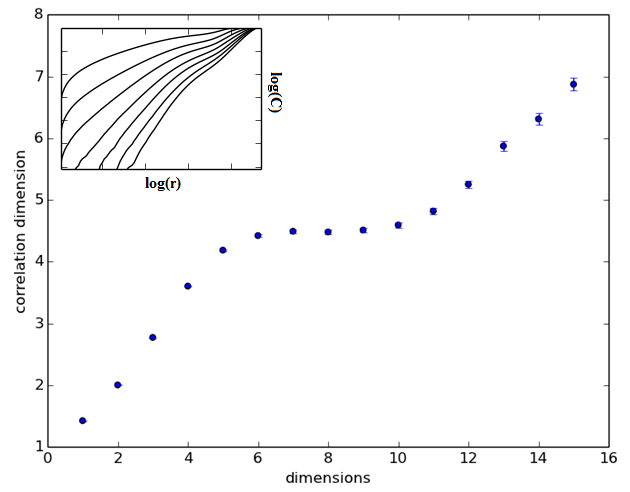}%
\caption{\label{fig:Cor_Dim} The slopes of the extracted linear region as embedding dimension is varied.  Inset displays the log of the correlation sum against the log of hypersphere radius averaged over 1000 reference vectors for embedding dimensions 1 to 15 (only odd dimensions shown).}
\end{figure}

In FIG \ref{fig:Cor_Dim}, we observe a plateau in the correlation dimension, which occurs at a value between 4 and 5. The plateau does not persist beyond 10 embedding dimensions.  We attribute this feature to the finite recording length, as the phase space becomes sparsely occupied with increasing embedding dimension.  A correlation dimension between 4 and 5 is consistent with our previous results, based on the false nearest neighbor test, which indicate that hair bundle dynamics contain between 3 and 6 degrees of freedom.

\subsection{Stimulus Near the Resonance Frequency}

\begin{figure}[h!]
\includegraphics[width=85mm]{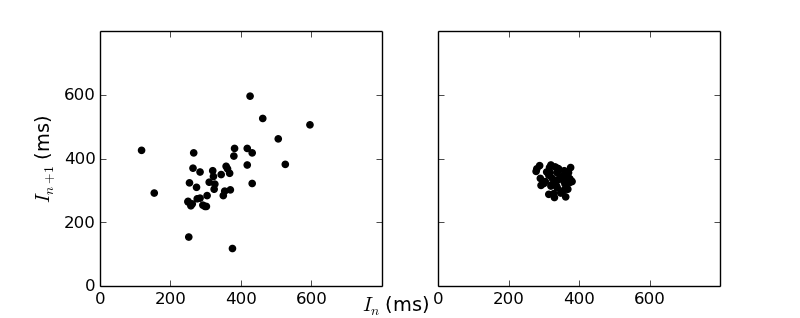}%
\caption{\label{fig:poincare_on_res} Poincar\'e maps representing spontaneous oscillations of hair bundles (left) and hair bundle response to a sinusoidal stimulus of $\sim$5 pN applied near the resonance frequency (right).}
\end{figure}

\begin{figure}[H]
\includegraphics[width=85mm]{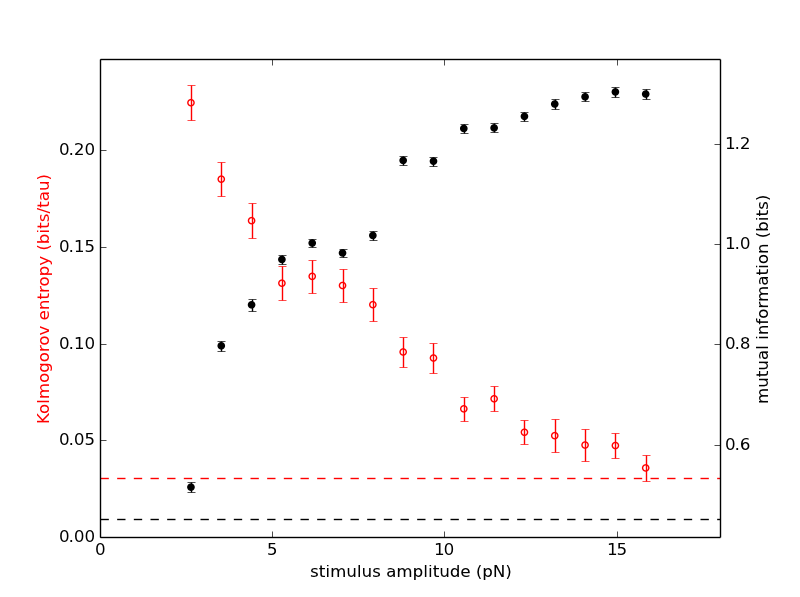}%
\caption{\label{fig:KE} The Kolmogorov entropy of an oscillatory hair bundle subject to a sinusoidal drive (open-red) and mutual information between the bundle and the stimulus (filled-black), plotted as functions of the stimulus amplitude.  Noise floors are indicated by horizontal dashed lines.  Stimulus frequency was selected to be near the resonance frequency of the hair bundle.}
\end{figure}

When hair bundles are stimulated near resonance, the Poincar\'e maps do not exhibit quasiperiodic transitions. Instead, as the forcing amplitude increases, the cloud in the $I_n$-$I_{n+1}$ plane shrinks in size and collapses onto a point (FIG \ref{fig:poincare_on_res}).  K-entropy decreases with increasing amplitude of the drive, while information transmission increases (FIG \ref{fig:KE}), comparable to the dependence observed with below-resonance stimulus.  As expected, the threshold for detecting weak stimulus is lower for the on-resonance case.

\subsection{Stimulus Above the Resonance Frequency} 

\begin{figure}[H]
\includegraphics[width=85mm]{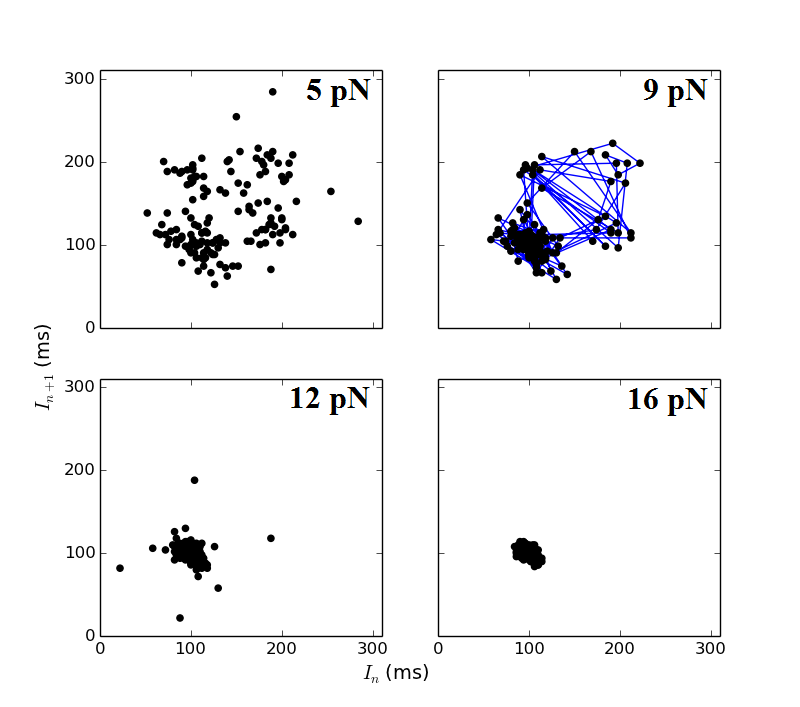}%
\caption{\label{fig:Poincare} Poincar\'e maps representing oscillations of hair bundles driven by a sinusoidal stimulus above the resonance frequency ($\omega \sim \frac{5}{3} \omega_0$).  The four clusters of points indicate a flicker between 1:1 and 1:2 mode-locking.  Blue lines connect consecutive points in the series.}
\end{figure}

A different type of transition from chaos to order is observed when the stimulus is applied at frequencies above resonance.  Rather than displaying a ring, the points on the Poincar\'e maps cluster into regions corresponding to integer multiples of the resonance frequency, indicative of high-order mode-locking (FIG \ref{fig:Poincare}). Rather than migrating around the edge (as in the case of a quasiperiodic transition), consecutive points cross over the center. Upon higher amplitudes of the drive, the cluster corresponding to 1:1 mode-locking dominates, and other clusters vanish.   K-entropy decreases with increasing amplitude of the drive, while information transmission increases (FIG \ref{fig:KE_above_res}), comparable to the dependence observed with below-resonance stimulus.  However, K-entropy initially increases for weak forcing.  We attribute this initial increase in K-entropy to flicker between 1:1 and 2:1 modes in the phase-locking of the bundle to the stimulus. 

\begin{figure}[H]
\includegraphics[width=85mm]{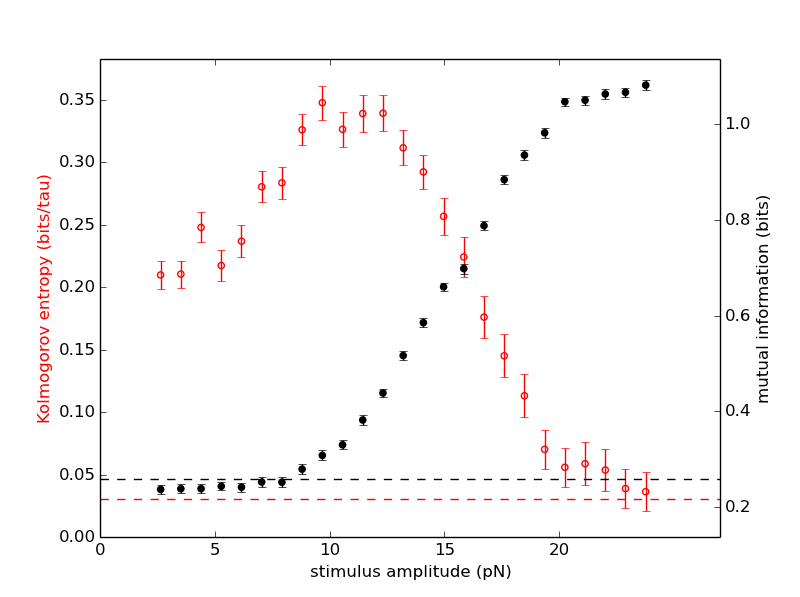}%
\caption{\label{fig:KE_above_res} The Kolmogorov entropy of an oscillatory hair bundle subject to a sinusoidal drive (open-red) and mutual information between the bundle and the stimulus (filled-black), plotted as a function of the stimulus amplitude.  Noise floors are indicated by horizontal dashed lines.  Stimulus frequency was above the resonance frequency of the hair bundle ($\omega \sim \frac{5}{3} \omega_0$).}
\end{figure}

\subsection{Theoretical Model} 
A sinusoidal stimulus of linearly increasing amplitude was applied to the theoretical model.  With $\alpha>0$, the model exhibits a quasiperiodic transition from chaos to order as forcing amplitude is increased.  The reduction in Komogorov Entropy and increase in mutual information are qualitatively the similar to those observed experimentally. Subtle differences are apparent, however; the model shows a more rapid increase in mutual information, implying that entertainment occurs more suddenly.

\begin{figure}[H]
\includegraphics[width=85mm]{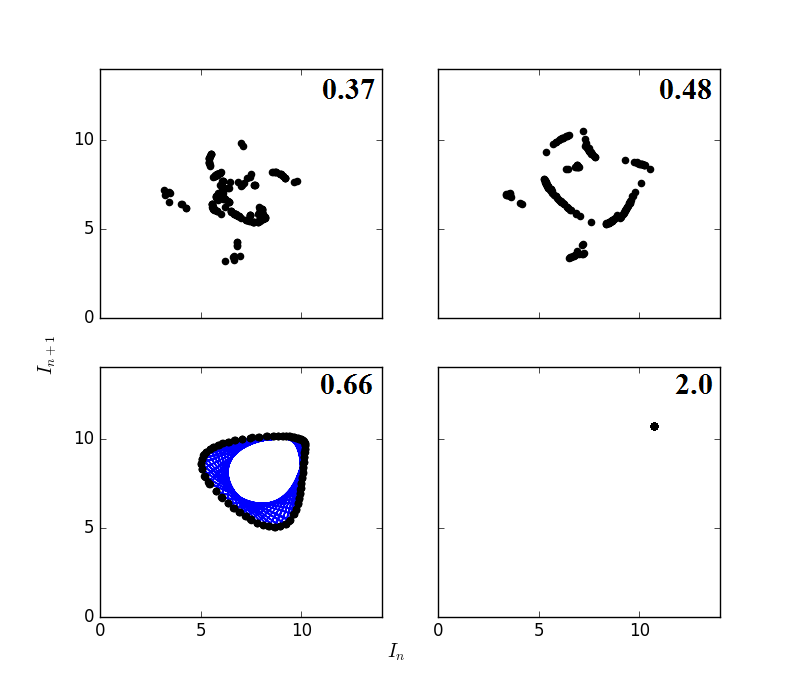}%
\caption{\label{fig:poincare_model} (Model) Poincar\'e maps representing oscillations of hair bundles driven by a sinusoidal stimulus below the resonance frequency ($\omega \sim \frac{2}{3} \omega_0$), for a range of forcing amplitudes, as indicated in the top right corners ($\alpha=0.15$).  Blue lines connect consecutive points in the series.}
\end{figure}

\begin{figure}[H]
\includegraphics[width=85mm]{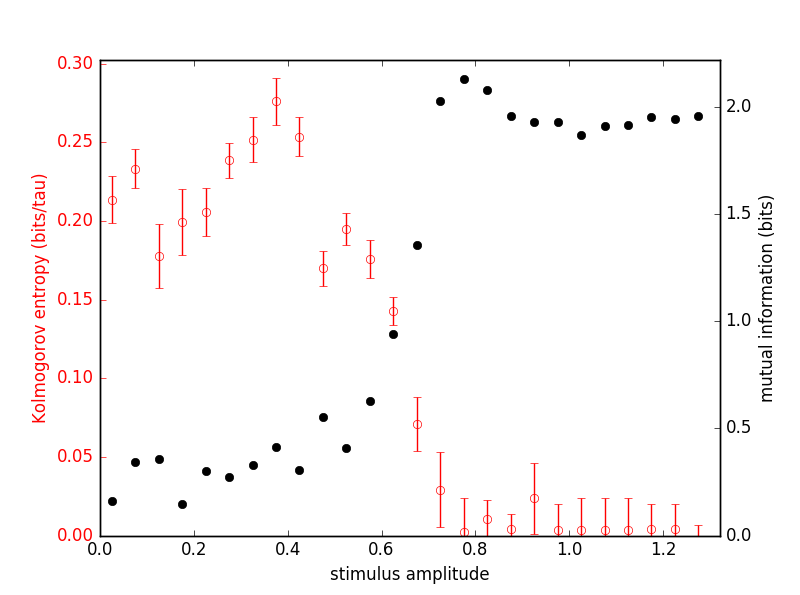}%
\caption{\label{fig:KE_above_res} (Model) The Kolmogorov entropy of an oscillatory hair bundle subject to a sinusoidal drive (red) and mutual information between the bundle and the stimulus (black), plotted as a function of the stimulus amplitude.  Stimulus frequency was below the resonance frequency of the hair bundle ($\omega \sim \frac{2}{3} \omega_0$).}
\end{figure}

\end{document}